\documentclass{iau}
\usepackage{graphicx}
\usepackage{natbib}
\title[JD 11.~~Anomalies] 
{Considerations in the Interpretation of Cosmological Anomalies}

\author[Hiranya V. Peiris]   
{Hiranya V. Peiris$^1$}

\affiliation{$^1$Department of Physics and Astronomy, University College London, \\ Gower Street, London WC1E 6BT, United Kingdom \\ email: {\tt h.peiris@ucl.ac.uk}}

\pubyear{2014}
\volume{306}  
\pagerange{119--126}
\jname{Statistical Challenges in 21st Century Cosmology}
\editors{A.F. Heavens, J.-L. Starck, A. Krone-Martins eds.}
\begin{document}

\maketitle

\begin{abstract}
Anomalies drive scientific discovery -- they are associated with the cutting edge of the research frontier, and thus typically exploit data in the low signal-to-noise regime. In astronomy, the prevalence of systematics --- both ``known unknowns'' and ``unknown unknowns'' --- combined with increasingly large datasets, the widespread use of {\it ad hoc} estimators for anomaly detection, and the ``look-elsewhere'' effect, can lead to spurious false detections. In this informal note, I argue that anomaly detection leading to discoveries of new physics requires a combination of physical understanding, careful experimental design to avoid confirmation bias, and self-consistent statistical methods. These points are illustrated with several concrete examples from cosmology.  


\end{abstract}

\firstsection 
\section{Introduction}

In the next decade, Big Data will form a pivotal part of the experimental landscape in cosmology, with a multitude of large surveys, from cosmic microwave background (CMB) experiments, photometric and spectroscopic galaxy surveys, 21 cm arrays, and direct-detection gravitational wave observatories, expected to yield a deluge of information about the origin and evolution of the Universe. The associated high data rates and huge data volumes, bringing with them the curse of dimensionality, pose challenging problems for data analysis and scientific discovery. Some of the challenges include: (1) data compression ({\it i.e.,} the formulation of almost-sufficient statistics), filtering, sampling, associated with very large datasets; (2) making robust and accurate inferences or conclusions from such datasets; (3) the small signal-to-noise regime in which discoveries at the research frontier are often made; (4) a very large model space; (5) cosmic variance ({\it i.e.,} the fact that we have access to a single realisation of an inherently stochastic cosmological model). 

We can imagine two qualitatively different kinds of modelling that would be required to extract the maximum information from this cornucopia of cosmological data. {\bf Mechanistic} (physical) modelling is what has driven cosmological discovery thus far -- in this case, forward modelling based on physics is feasible, leading directly to standard parameter estimation and model comparison analyses that form the bread and butter of modern cosmology. However, as cosmological data probe deeper into the non-linear regime of the evolution of structure, against complex foregrounds, {\bf empirical} (data-driven) modelling --- characterising relationships discovered in the data --- will gain increasing prominence. Such models may be purely data-driven, or qualitatively based on physics but be required in cases where forward modelling is infeasible; they may be used to postulate new theories, or generate statistical predictions for new observables.

In this context, the treatment of data anomalies acquires great importance. Anomalies --- unusual data configurations --- can consist of statistical outliers, unusual concentrations of data points or sudden behaviour changes. They may arise from chance configurations due to random fluctuations, systematics (unmodelled astrophysics; instrument/detector artefacts; data processing distortions), or they can indicate genuinely new discoveries. Determining into which of these categories a given anomaly falls is fraught with difficulty, especially as humans have evolved in such a way that we can often spuriously identify patterns in data where none exist, a phenomenon known as {\it pareidolia}. When considering whether a cosmological anomaly indicates new physics, one must consider that many such anomalies are often identified using {\it a posteriori} estimators, which leads to spurious enhancements of detection significance. In the absence of an alternative model for comparison, one often cannot account for the ``look-elsewhere effect'' arising from multiple testing, or formulate model priors to compare with the standard model. In the absence of an alternative theory, how can we judge if a given anomaly represents new physics? In the rest of this article, I will summarise several case studies demonstrating different approaches to answering this question.

\section{Assessing anomalies}

There are two distinct steps associated with anomalies in the cosmological context. The first involves finding data anomalies in the first place --- this provides an area of rich algorithmic development related to assessing measures of irregularity, unexpectedness, unusualness etc., especially in the coming ``Big Data'' era. The second involves drawing inferences from the results of the search for anomalies --- {\it i.e.,} assessing whether the anomaly is due to random chance, or whether it represents an unknown mechanism which may point to new physics (or, more prosaically, systematics). In making these inferences, the particle physicists' ``look-elsewhere effect'', or multiple testing, comes into play. One must correct for any {\it a posteriori} choices that were made in the process of detecting an anomaly, and properly account for the possible ways in which an anomaly could have shown up (but did not). 

In cases where an alternative model motivated the search for an anomaly or the formulation of the statistic in which an anomaly was discovered, the probability that the anomaly represents a new physical mechanism (versus random chance) can be assessed using Bayesian model comparison: the model and parameter priors account for the uncertainties associated with multiple-testing. Even in this case, however, encapsulating all the relevant uncertainties can be quite difficult. For example, \cite{Cruz:2007pe} carried out a Bayesian model comparison analysis of the well-known cosmic microwave background (CMB) anomaly termed the {\it Cold Spot} and concluded that it was likely a texture (a type of spatially-localised cosmic defect). This was a highly sophisticated analysis --- one of the first principled Bayesian model comparison analyses in the context of CMB anomalies. Nevertheless, it represents an incomplete attempt to account for {\it a posteriori} selection effects associated with basing the analysis on a single feature at a particular location which is known to be a statistical outlier. 

The texture model predicts a statistically isotropic distribution of spatially-localised features in the CMB sky (with an expectation value for the sky fraction covered by textures), and textures can lead to both hot and cold spots in the CMB. Interpreted as a correction for the ``look-elsewhere'' effect of having analysed a patch of sky containing the Cold Spot, the \cite{Cruz:2007pe} analysis accounts for the expected sky fraction covered by textures in a patch, but does not account for the fact that textures could be placed anywhere on the sky. The coordinates of the central position of the texture template should be marginalised over in the context of statistical isotropy, whereas the texture template was centred on the Cold Spot in the analysis. There is also a factor of two associated with the fact that the feature could have been a hot spot, rather than a cold spot. In \cite{Feeney:2012jf, Feeney:2012hj}, where the texture hypothesis is formulated in terms of a hierarchical Bayesian model, the impact of these {\it a posteriori} selection effects becomes clear and, when fully accounting for these uncertainties in computing the marginal likelihood, there is no preference for augmenting the standard cosmological model with textures. 

This example highlights how the ``look-elsewhere'' effect can spuriously enhance the significance of the anomaly when all the relevant uncertainties are not incorporated into the formulation of an alternative model representing new physics. Nevertheless, if one possesses alternative model(s) with well-motivated priors, the standard Bayesian model comparison framework can be brought to bear on the problem, providing the most straightforward scenario for assessing anomalies.

\section{``Just-so'' models}

If the anomaly is detected as an unusual property of a large dataset and is not motivated by an alternative model, accounting for the ``look-elsewhere'' effect can be highly non-trivial. In this context, designer theories that stand in for best-possible explanations --- ``just-so'' models --- can prove useful in terms of gaining an intuition for whether the anomaly provides evidence for new physics. 

In this case one would proceed as follows: (a) find a {\it designer theory} or ``just-so'' model which maximises the likelihood of the anomaly; (b) thus determine the maximum available likelihood gain for this particular anomaly with respect to the standard model, or null hypothesis; (c) judge whether this is compelling given the baroqueness of the designer theory. 

Such an example can be found in the context of another well-known CMB anomaly often termed the $C(\theta)$ anomaly \citep{Copi:2006tu}. It is a long-standing observation that the statistic $S_{1/2} = \int_{60^\circ}^{180^\circ} [C(\theta)]^2 \cos(\theta) d\theta$, where $C(\theta)$ is the angular correlation function of the CMB, is anomalous when evaluated outside typical Galactic sky cuts. Here, the lower integration limit is an {\it a posteriori} choice \citep{Bennett2011}. Under the standard $\Lambda$CDM model, the probability of obtaining a cut-sky $S_{1/2}$ statistic of the observed value or less is $\sim 0.03$\%; however, it is notable that the {\it full sky} value of $S_{1/2}$ (evaluated on reconstructed full sky maps or Galactic foreground-cleaned maps) is not anomalous. 

\begin{figure}[htbp]
    \centering
       \includegraphics[width=.55\textwidth]{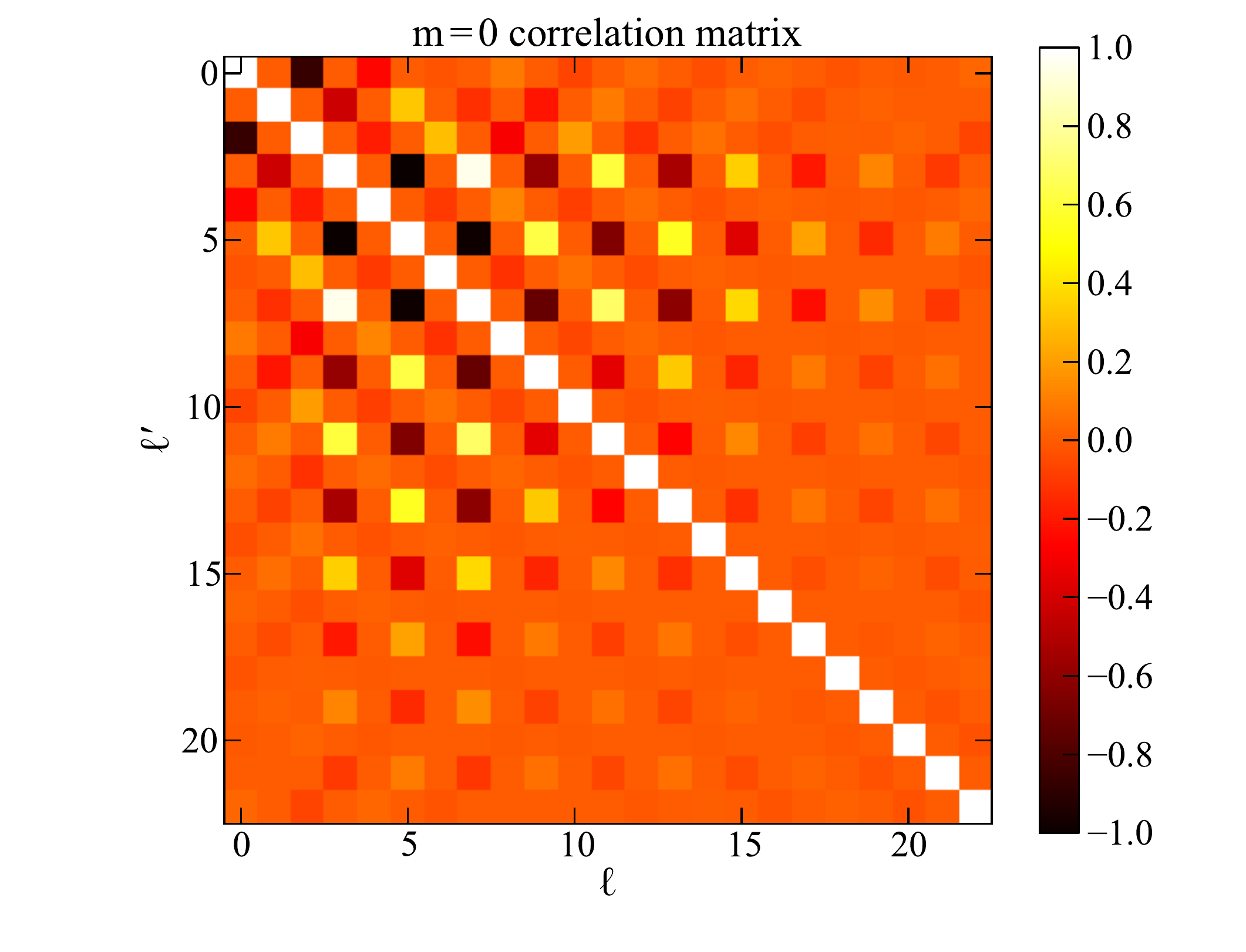}
    \caption{The $m=0$ component of the covariance matrix defining the designer anisotropic model derived for the $S_{1/2}$ anomaly \citep{PP2010}.}
    \label{fig:designer}
\end{figure}

It is possible that such an observation can arise in a cosmological model which breaks statistical isotropy, in contrast to the standard $\Lambda$CDM model; however, a specific well-motivated broken-isotropy model predicting the observed characteristics of the large-angle CMB sky is not currently available. \cite{PP2010} used a convex optimisation algorithm to maximise the likelihood of the cut sky $S_{1/2}$ statistic subject to fixed full sky angular power spectrum multipoles $C_\ell$ over all anisotropic Gaussian models with zero mean\footnote{The covariance matrix of spherical harmonic coefficients $a_{\ell m}$ can be can be arbitrarily correlated, as long as it is positive-definite.}. Using this technique, they constructed a designer anisotropic model  (Fig.~\ref{fig:designer}) which gives the maximum likelihood improvement over $\Lambda$CDM: this model, which had $\sim 6900$ degrees of freedom (compared to the isotropic case with eight) gave an improvement in likelihood of $\ln {\cal L} \sim 5$. This allowed for a finite limit to be placed on the Bayesian statistical gain available under a wide class of alternative straw-man models, providing a plausible way to probe the significance of this {\it a posteriori} anomaly.

It is interesting to note that the decision-making process after computing the likelihood gain supplied by the just-so model involves a subjective judgement. For example, \cite{PP2010} concluded that the level of fine-tuning involved in this designer model in order to provide a relatively small likelihood gain did not make for a compelling pointer towards new physics. However, a more speculative scientist may make a different decision to continue the search for an alternative theory, assuming that the number of degrees of freedom can be dramatically reduced by unknown symmetries. In this case, the designer model may provide physical intuition about the correlation properties necessary in such a theory in order to reproduce the observations. 

A further example of the utility of just-so models in cosmology where physically-motivated model priors are not available is found in the context of the number of relativistic degrees of freedom $N_\mathrm{eff}$, which may differ from the standard value of $3.046$ due to, {\it e.g.}, extra neutrino species. The fundamental debate here focuses on whether tensions observed in cosmological datasets require the standard cosmological model to be augmented with a non-standard $N_\mathrm{eff}$. While this may appear to be a straightforward Bayesian model comparison problem, the model space leading to $N_\mathrm{eff}$ differing from the standard value is vast, and simple priors on $N_\mathrm{eff}$ may not capture the physical prior uncertainties. In this context, \cite{Verde2013} show that the {\it profile likelihood ratio} \citep{wilks1938} --- the ratio of the maximum likelihood conditioned on a particular value of the parameter of interest and the overall, unconditional maximum likelihood --- provides the parameter value at which a designer model would maximise the Bayesian evidence. While it is unreasonable to assess a model by tuning it to the same data being used to test it, this upper limit on the marginal likelihood provides an heuristic assessment of the possible need for an extra parameter in the absence of well-motivated parameter priors, as well as a pointer towards its most likely value.

\section{Data-driven models}

As mentioned above, data-driven models are often constructed {\it a posteriori} to fit various statistical anomalies observed in large data sets. In order to be testable, such tuned models must, of necessity, make predictions for new data beyond those they were designed to fit. In the context of CMB anomalies, several studies demonstrate how models designed to mimic apparent isotropy-breaking features in the CMB temperature field and features in the CMB temperature power spectrum can be tested with new data not used in constructing the model, such CMB polarisation \citep{Dvorkin:2007jp, Mortonson:2009qv}, large-scale structure data \citep{PullenHirata}, and non-Gaussianity \citep{Adshead:2011bw,Peiris:2013opa}.

The statistical technique of cross-validation can be used as a powerful method to test the consistency of physical inferences from datasets by constructing a data-driven model describing part of the data, and checking how well it predicts the rest of the data. For example, this technique has been used to characterize the deviation from scale invariance in the primordial power spectrum in a minimally-parametric way \citep{Verde:2008zza,Peiris:2009wp,Bird:2010mp}. The idea is as follows: (1) Choose a functional form which allows a great deal of freedom in the form of the deviation from scale invariance ({\it e.g.}, smoothing splines)\footnote{Naively fitting this to the data will lead one to fit the fluctuations due to noise, with arbitrary improvement in the goodness-of-fit due to over-fitting.}. (2) Now perform cross-validation: set aside some of the data (validation set), fit the rest (training set), and see how well it predicts the validation set. A very good fit to the training set, which poorly predicts the validation set, indicates over-fitting of noisy data. (3) The final ingredient is a parameter that penalises fine-scale structure of the functional form. By performing cross-validation as a function of this parameter, one can judge when the structure in the smoothing spline is what the data requires without fitting the noise. The technique is also very powerful in detecting systematic issues in data analysis: \cite{Verde:2008zza} identified a ``kink" in the reconstructed power spectrum from WMAP at a particular scale which subsequently turned out to be a problem in point source subtraction \citep{2008ApJ...688....1H, Peiris:2009wp}. 

\section{Blind analysis}
The data analysis challenges of next-generation cosmological surveys require careful experimental design to minimize false detections due to experimenters' (subconscious) bias. This has been an integral part of the particle physics and medical research culture for decades, but has yet to find broad adoption within cosmology. As cosmological constraints make the transition from {\it precise} to {\it accurate}, and the search for new physics leads to the hunt for small signals embedded within the $\Lambda$CDM cosmology, the systematic bias due to non-blind experimental design can no longer be ignored if we hope to make convincing claims for paradigm-shifting new physics from cosmological data. 

Blind analysis is based on the simple idea that the {\it value} of a measurement does not contain any information about its {\it correctness}. Knowing the value of a measurement is therefore of no use in performing the analysis itself. Blind analysis is necessary because data collection, analysis and inference necessarily involves a human stage, which can lead to unquantifiable subjective inferences. Some examples of the origin of such bias include: looking for bugs when a result does not conform to expectation (and not looking for them when it does); looking for additional sources of systematic uncertainty when a result does not conform to expectation; deciding whether to publish, or wait for more data; choosing cuts while looking at the data and knowing whether it fits expectations; preferentially keeping / dropping outlier data. Examples of good experimental design which are used in cosmology, such as ``double unblind'' (doing two independent analyses in parallel), ``mock data'' analysis, and ``semi-blind'' (using a fraction of the data to calibrate the analysis and freezing the pipeline before doing the final analysis), while extremely helpful in their own right, nevertheless do not constitute blind analysis. 

 \cite{Harrison2002} and \cite{2003sppp.conf..166R} give examples of systematic trends in measurements (periods of surprisingly small variation, followed by jumps of several standard deviations) in the presence of experimenters' bias using examples from particle physics, and provide some excellent strategies for implementing blind analysis, including:
\begin{itemize}
\item{{\bf Encrypting the science result.} {\it e.g.}, adding a non-changing random number (not revealed to the analyst) to a numerical result or transforming a variable, in order to thwart preconceptions due to ``expected'' results. It is important to note that it is not necessary to blind how the result {\it changes} due to changes in the analysis pipeline, or to blind calibration data.}
\item{{\bf Hiding the ``signal region''.} This is useful when the observable is a peak or localised feature in datasets.}
\item{{\bf Blind injection of signals into the data.} This is useful to test biases in rare event searches, and has been adopted by the gravitational wave direct detection community.}
\item{{\bf Mixing in an unknown fraction of simulated data} during calibration, etc.}
\end{itemize}

The first of these, the so-called {\em hidden offset} method, is a straightforward technique to implement at the level of the parameter estimation step for essentially all cosmological data types; it works as follows. The parameter estimation code adds a fixed, unknown random number to the fitted value of the measured parameter, $x^* = x + {\cal R}$; $x^*$ is returned with the true error and the likelihood value instead of $x$. ${\cal R}$ can be set {\it e.g.}, by sampling from a Gaussian of mean zero and standard deviation $\sim$ few times the experimental standard deviation. Relative changes in the result as the analysis changes can be hidden using a second offset. Informative visual aspects in plots may need to be hidden as well. 

While blind analysis within observational cosmology may be conceptually more difficult than in an experimental field such as particle physics, nevertheless numerous considerations motivate us to adopt at least some practices that lead to the mitigation of experimenters' bias. Such considerations include: (i) the accuracy and reliability of scientific inferences from next-generation surveys; (ii) best return on the enormous investment of public funds in cosmological surveys, including satellite experiments and large data analysis efforts requiring huge teams, which may be very difficult --- if not impossible --- to replicate due to cost considerations; (iii) the substantial wasted scientific effort in going down blind alleys due to premature announcements of false detections, a side effect of which is the potential damage to the public perception of science; and (iv) missing discoveries of new physics due to overly-cautious treatment of data that do not agree with expectations. 

Even simply thinking about how to blind an analysis pipeline can lead to a greater understanding of potential pitfalls. Adopting blind analysis practices requires a shift in community standards, and leadership by large data collaborations. A very welcome shift in this direction has been made by the weak lensing community, {\it e.g.}, in the recent CFHTLenS analyses \citep{2012MNRAS.427..146H, Fu:2014loa}. The POLARBEAR $B$-mode polarisation blind analysis \citep{Ade:2014afa} represents a milestone in the CMB context, and such techniques have been in use for some time in supernova cosmology \citep{Conley:2006qb}. In general, setting all the free parameters in analysis pipelines using only null-tests is a good blind analysis practice. An example of blind mitigation of systematics in constraining primordial non-Gaussianity using quasar surveys \citep{2014MNRAS.444....2L,2014arXiv1405.4315L} is presented elsewhere in these proceedings. 

\section{Summary}
I have presented an informal summary, from a practitioner's viewpoint, of techniques for evaluating cosmological anomalies. The discussion has emphasised the pitfalls associated with multiple-testing, or the ``look-elsewhere'' effect. I have given practical examples of techniques that can be used in the absence of alternative models in order to gain intuition on whether or not a given cosmological anomaly represents new physics. Finally, I have discussed blind analysis as a strategy to guard against experimenters' subconscious bias, which introduces ``unknown unknowns'' into physical inferences. These techniques do not represent a cure-all for data analysis problems, but rather cultivate a mindset that attempts to rise to Feynman's challenge: {\it ``The first principle is that you must not fool yourself --- and you are the easiest person to fool.''}
\\

{\it Acknowledgements}--- I thank Stephen Feeney, David Hand, Boris Leistedt, Daniel Mortlock, Andrew Pontzen, Aaron Roodman, and Licia Verde for influencing my evolving take on this topic. I am grateful to Andrew Pontzen for generating Fig.~\ref{fig:designer}, which was previously unpublished.

\bibliography{references}

\begin{thebibliography}{25}
\expandafter\ifx\csname natexlab\endcsname\relax\def\natexlab#1{#1}\fi

\bibitem[{Ade {et~al.}(2014)}]{Ade:2014afa}
Ade, P. {et~al.} 2014, ApJ, 794, 171

\bibitem[{Adshead {et~al.}(2011)Adshead, Hu, Dvorkin, \&
  Peiris}]{Adshead:2011bw}
Adshead, P., Hu, W., Dvorkin, C., \& Peiris, H.~V. 2011, Phys.Rev., D84, 043519

\bibitem[{{Bennett} {et~al.}(2011){Bennett}, {Hill}, {Hinshaw}, {Larson},
  {Smith}, {Dunkley}, {Gold}, {Halpern}, {Jarosik}, {Kogut}, {Komatsu},
  {Limon}, {Meyer}, {Nolta}, {Odegard}, {Page}, {Spergel}, {Tucker}, {Weiland},
  {Wollack}, \& {Wright}}]{Bennett2011}
{Bennett}, C.~L., {Hill}, R.~S., {Hinshaw}, G., {et~al.} 2011, ApJS, 192, 17

\bibitem[{Bird {et~al.}(2011)Bird, Peiris, Viel, \& Verde}]{Bird:2010mp}
Bird, S., Peiris, H.~V., Viel, M., \& Verde, L. 2011, MNRAS, 413, 1717

\bibitem[{Conley {et~al.}(2006)}]{Conley:2006qb}
Conley, A.~J. {et~al.} 2006, ApJ, 644, 1

\bibitem[{Copi {et~al.}(2007)Copi, Huterer, Schwarz, \& Starkman}]{Copi:2006tu}
Copi, C., Huterer, D., Schwarz, D., \& Starkman, G. 2007, Phys.Rev., D75,
  023507

\bibitem[{Cruz {et~al.}(2007)Cruz, Turok, Vielva, Martinez-Gonzalez, \&
  Hobson}]{Cruz:2007pe}
Cruz, M., Turok, N., Vielva, P., Martinez-Gonzalez, E., \& Hobson, M. 2007,
  Science, 318, 1612

\bibitem[{Dvorkin {et~al.}(2008)Dvorkin, Peiris, \& Hu}]{Dvorkin:2007jp}
Dvorkin, C., Peiris, H.~V., \& Hu, W. 2008, Phys.Rev., D77, 063008

\bibitem[{Feeney {et~al.}(2013)Feeney, Johnson, McEwen, Mortlock, \&
  Peiris}]{Feeney:2012hj}
Feeney, S.~M., Johnson, M.~C., McEwen, J.~D., Mortlock, D.~J., \& Peiris, H.~V.
  2013, Phys.Rev., D88, 043012

\bibitem[{Feeney {et~al.}(2012)Feeney, Johnson, Mortlock, \&
  Peiris}]{Feeney:2012jf}
Feeney, S.~M., Johnson, M.~C., Mortlock, D.~J., \& Peiris, H.~V. 2012,
  Phys.Rev.Lett., 108, 241301

\bibitem[{Fu {et~al.}(2014)Fu, Kilbinger, Erben, Heymans, Hildebrandt,
  {et~al.}}]{Fu:2014loa}
Fu, L., Kilbinger, M., Erben, T., {et~al.} 2014

\bibitem[{Harrison(2002)}]{Harrison2002}
Harrison, P.~F. 2002, Journal of Physics G: Nuclear and Particle Physics, 28,
  2679

\bibitem[{{Heymans} {et~al.}(2012){Heymans}, {Van Waerbeke}, {Miller}, {Erben},
  {Hildebrandt}, {Hoekstra}, {Kitching}, {Mellier}, {Simon}, {Bonnett},
  {Coupon}, {Fu}, {Harnois D{\'e}raps}, {Hudson}, {Kilbinger}, {Kuijken},
  {Rowe}, {Schrabback}, {Semboloni}, {van Uitert}, {Vafaei}, \&
  {Velander}}]{2012MNRAS.427..146H}
{Heymans}, C., {Van Waerbeke}, L., {Miller}, L., {et~al.} 2012, MNRAS, 427, 146

\bibitem[{{Huffenberger} {et~al.}(2008){Huffenberger}, {Eriksen}, {Hansen},
  {Banday}, \& {G{\'o}rski}}]{2008ApJ...688....1H}
{Huffenberger}, K.~M., {Eriksen}, H.~K., {Hansen}, F.~K., {Banday}, A.~J., \&
  {G{\'o}rski}, K.~M. 2008, ApJ, 688, 1

\bibitem[{{Leistedt} \& {Peiris}(2014)}]{2014MNRAS.444....2L}
{Leistedt}, B. \& {Peiris}, H.~V. 2014, MNRAS, 444, 2

\bibitem[{{Leistedt} {et~al.}(2014){Leistedt}, {Peiris}, \&
  {Roth}}]{2014arXiv1405.4315L}
{Leistedt}, B., {Peiris}, H.~V., \& {Roth}, N. 2014, ArXiv e-prints

\bibitem[{Mortonson {et~al.}(2009)Mortonson, Dvorkin, Peiris, \&
  Hu}]{Mortonson:2009qv}
Mortonson, M.~J., Dvorkin, C., Peiris, H.~V., \& Hu, W. 2009, Phys.Rev., D79,
  103519

\bibitem[{Peiris {et~al.}(2013)Peiris, Easther, \& Flauger}]{Peiris:2013opa}
Peiris, H., Easther, R., \& Flauger, R. 2013, JCAP, 1309, 018

\bibitem[{Peiris \& Verde(2010)}]{Peiris:2009wp}
Peiris, H.~V. \& Verde, L. 2010, Phys.Rev., D81, 021302

\bibitem[{{Pontzen} \& {Peiris}(2010)}]{PP2010}
{Pontzen}, A. \& {Peiris}, H.~V. 2010, Phys. Rev., D81, 103008

\bibitem[{{Pullen} \& {Hirata}(2010)}]{PullenHirata}
{Pullen}, A.~R. \& {Hirata}, C.~M. 2010, JCAP, 5, 27

\bibitem[{{Roodman}(2003)}]{2003sppp.conf..166R}
{Roodman}, A. 2003, in Statistical Problems in Particle Physics, Astrophysics,
  and Cosmology, ed. L.~{Lyons}, R.~{Mount}, \& R.~{Reitmeyer}, 166

\bibitem[{{Verde} {et~al.}(2013){Verde}, {Feeney}, {Mortlock}, \&
  {Peiris}}]{Verde2013}
{Verde}, L., {Feeney}, S.~M., {Mortlock}, D.~J., \& {Peiris}, H.~V. 2013, JCAP,
  9, 13

\bibitem[{Verde \& Peiris(2008)}]{Verde:2008zza}
Verde, L. \& Peiris, H.~V. 2008, JCAP, 0807, 009

\bibitem[{Wilks(1938)}]{wilks1938}
Wilks, S.~S. 1938, The Annals of Mathematical Statistics, 9, 60

\end{thebibliography}

\end{document}